\pdfoutput=1
\documentclass[aps,pra,reprint,showpacs]{revtex4-1}
\usepackage{graphicx}
\usepackage{amsmath}
\usepackage{amssymb}
\usepackage{amsfonts}
\usepackage{dcolumn}
\usepackage{dsfont}
\usepackage{latexsym}
\usepackage{rotating}
\usepackage{color}
\usepackage{latexsym}
\usepackage{bbm}
\usepackage{subfigure}
\usepackage{float}
\usepackage{epsfig}
\usepackage{psfrag}
\usepackage{natbib}
\usepackage{bm}
\usepackage{amsthm}
\usepackage{eucal}
\usepackage{mathrsfs}
\usepackage{url}
\usepackage{braket}
\usepackage{mathtools}
\usepackage{amsmath,amssymb}
\usepackage{epsfig}
\usepackage{amsmath,amssymb}
\usepackage{hyperref}

\usepackage{color}

\usepackage{hyperref}
\hypersetup{
colorlinks=true,final=true,
        linkcolor=blue,
        citecolor=blue,
        filecolor=blue,
        urlcolor=blue,
}

\begin{document}

\title{Dynamics and selective temporal focusing of a time truncated Airy pulse in varying dispersive medium }

\author{Aritra Banerjee$^{\star,1}$ and Samudra Roy$^{\dagger,1,2}$}
\affiliation{$^1$Department of Physics, Indian Institute of Technology Kharagpur, W.B. 721302, India}
\affiliation{$^2$Centre for Theoretical Studies, Indian Institute of Technology Kharagpur, W.B. 721302, India}
\email{$^\dagger$samudra.roy@phy.iitkgp.ac.in\\
$^\star$aritra@iitkgp.ac.in}

\begin{abstract}
We theoretically investigate the dynamics of a time truncated Airy pulse under longitudinally varying dispersion. The realistic waveguide geometry is proposed that offers linear or oscillating dispersion profile. By solving the dispersion equation, we theoretically investigate how a linear variation of the group-velocity dispersion (GVD) over space affects the parabolic trajectory of an accelerating finite energy airy pulse (FEAP). It is demonstrated that, suitable adjustment of GVD can lead to unusual quasi-linear trajectory of the accelerating airy pulse. The impact of the periodic GVD on airy dynamics is more interesting where FEAP exhibits oscillatory trajectory with periodic peak power modulation. We theoretically estimate optimised length of the waveguide delivering maximum power at the output by solving the transcendental relation between GVD modulation strength and period. The effect of oscillatory higher order dispersion is dramatic for optical airy pulse where it experiences singularity points during propagation. At singularity points airy pulse loses its identity and flips over in time. The rich dynamics of FEAP near singular point is carefully investigated by solving the propagation equation analytically. In this report we provide detail theoretical analysis to achieve selective temporal focusing of FEAP which may be useful for application. All theoretical predictions are verified numerically and  the agreement is found to be excellent.

\end{abstract}

\maketitle

 \section{Introduction}

\noindent   Airy function was first introduced as an accelerating undistorted solution of the time dependent Schr\"{o}dingers equation in free space \cite{Berry}. The realisation of the Airy function in the optical domain by Siviloglou et.al. \cite{Siviloglou} opened up new avenues in the study of accelerating optical beams and pulses.  After the experimental observation of the truncated finite energy version of the Airy function as a phase modulated gaussian beam, many works have been reported exploring the unique properties of the Airy beam like self-acceleration, quasi diffraction free and self healing nature \cite{Siviloglou,Siviloglou_b,Broky}. Exploiting the isomorphism between the spatial diffraction and temporal dispersion the temporal counterpart of the finite energy Airy beam is realised as a time truncated finite energy Airy pulse (FEAP) \cite{Saari}. Airy pulses are the waveforms which travel undistorted in linear dispersive mediums where the effect of higher order dispersions is negligible and it follows a parabolic trajectory in time. The trajectory of the pulse depends on the dispersion characteristics of the waveguide through which the pulse propagates. Though FEAP is not an exact solution of the dispersion equation but still it keeps the unique properties of the Airy function intact for a finite distance. After the discovery of the self healing  Airy pulse, several interesting works have been done in the temporal domain like  absolute focusing under third order dispersion (TOD) \cite {driben,Shaarawi}, soliton shedding from the high power Airy pulse  \cite{Fattal}, mimicking event horizon through Airy-soliton collision \cite {Yang}, generation of new frequency components by the collision of Airy-soliton \cite{Roy}, Supercontinuum generation \cite {Ament} etc. The description of Airy functions in time domain also opens up exciting applications ranging from bioimaging, nano-machining to plasma physics\cite{Courvoisier,Englert,Gotte,javier,sarpe,Thomas}.

The previous works mentioned above have been done mostly for longitidinally static chromatic dispersion parameters where the possibility of manipulating  the pulse shape and its trajectory is limited. In this work,we try to explore the properties of the FEAP under longitudinally varying group velocity dispersion (GVD) profile. We consider a linear as well as the periodic variation of GVD over space and try to investigate its consequence in Airy dynamics in time frame. The periodic modulation of the dispersion is common in optical fibers where the core diameter varies periodically with fiber length, such fibers are called \textit{dispersion oscillating fiber} (DOF) \cite{Biancalana,Droques,Finot,Mussot}. In DOF the optical Kerr nonlinearity is also weakly modulated which leads to additional modulation instability (MI) side-band pairs \cite{Mussot,Trillo}. In nonlinear domain the soliton dynamics also becomes interesting when dispersion oscillates periodically over waveguide length. The longitudinal oscillation of dispersion in fiber results controlled soliton fission \cite{Sysoliatin,Sysoliatin_b}and also leads to multiple quasi-phase matched dispersive waves \cite{Wright,Conforti} resulting tailor-made supercontinuum generation \cite{Hickstein}. Very recently the optical analogue of dynamical Casimir effect is observed in varying dispersion fiber \cite{Vezzoli}. When we find substantial seminal works on optical solitons,the study of Airy like pulse in longitudinally varying dispersion is to some extent limited. Very few attempts were made previously to understand the behaviour of the FEAP in the environment of oscillating GVD  \cite{Bai,Driben_b}. These studies are mainly based on numerical computation that may hinder few key characteristics beneath in the theoretical solution. The dynamics of FEAP is far more complicated in realistic domain and requires an extensive investigation.  

To capture the behaviour of the FEAP in realistic systems, we design waveguides with longitudinally varying dispersion profiles. Imposing linear and periodic geometry on Si-based waveguides we obtain the GVD that varies linearly or oscillates around an average value over distance. Exploiting the COMSOL simulation we demonstrate if the width of the waveguide has a linear variation with propagation distance, the GVD  becomes a linear function of distance. The usual ballistic temporal trajectory of the airy pulse is significantly manipulated by varying dispersion and  one can even get quasi-linear path. We obtain a complete analytical solution of the moving airy pulse in varying dispersion environment and explain the phenomenon with the support of numerical simulation. The dynamics of the FEAP becomes more complicated when it encounters oscillating GVD. Waveguides with periodically varying widths offer an oscillating dispersion which radically change the behaviour of airy pulse specially when TOD is non-vanishing. TOD leads to a singularity in the airy pulse solution and because of which the temporal distribution of the pulse flips \cite{driben}. Under periodic TOD one can witness multiple flipping of the waveform  that takes places at periodic intervals. At flipping points the airy pulse losses its characteristics and focus tightly in the neighbourhood of the flipping zone. The entire propagation dynamics of the FEAP under periodic TOD is investigated by solving the linear dispersion equation analytically in different zones. The set of solutions reveal that under oscillating TOD the pulse evolves through periodic focusing and one can achieve selective absolute focusing of the pulse by selecting suitable dispersion modulation factor. Absolute focusing is an unique phenomenon where entire energy of the airy pulse is confined tightly. In application point of view the selective focusing may be interesting as we can deliver the entire energy of the time truncated airy pulse at specific output.

\section{Dynamics of FEAP under linear GVD variation} 

\noindent The wave number $\beta(\omega)$ of an optical wave is in general a function of frequency ($\omega$) and can be expanded in a Taylor series around the carrier frequency ($\omega_0$) as, $\beta(\omega)=\beta_0+\beta_1(\omega-\omega_0)+\frac{1}{2}\beta_2(\omega-\omega_0)^2+...$, where $\beta_0=\beta(\omega_0)$ and $\beta_j (\omega)=\frac{d^j \beta (\omega)}{d \omega^j}|_{\omega=\omega_0}$  $(j=1,2,3,4..)$. The GVD $\beta_2 (\omega)$ is an intrinsic property of an optical waveguide and can be manipulated by tailoring the waveguide geometry. Si-based planar waveguides are found to be the ideal candidate in controlling the dispersion profile in an arbitrary way. For a linear GVD variation over space we can model the dispersion profile as $\beta_2(z)=\beta_{20}+gz$, where $\beta_{20}$ is the GVD parameter at the input and it depends on the launching wavelength of the pulse. The parameter $g$ determines the rate of change of $\beta_2$ with the propagation distance $z$. Under such dispersion profile the dynamics of a FEAP $u(\xi,\tau)$ can be modelled as \cite{Agarwal}, 
\begin{equation} \label{q1}
i\frac{\partial u}{\partial \xi}=\frac{{{\delta}_{2}(\xi)}}{2}\frac{{{\partial }^{2}}u}{\partial {{\tau}^{2}}}-i\widetilde{\alpha}\xi,
\end{equation}
\noindent where the parameters are normalised as,  $u=U/\sqrt{P_0}$, $\xi=z{L_D}^{-1}$, $\tau=(t-z{v_g}^{-1})/t_0=T/{t_0}$. $U$, $P_0$  and $v_g$ respectively represent the optical field, input peak power and group velocity in real unit.  The width of the main lobe of FEAP is defined by $t_0$ which we consider $\sim 100$ fs.  $z$ and $t$ represent the space and time variables with physical units. The linear loss is normalised as $\widetilde{\alpha}=\alpha L_D$. For Si-based waveguide $\alpha \sim $ 1 dB/cm \cite{zou,Mashanovich}. The dispersion length $L_D$ is defined as $L_D=t_0^2/|\beta_{20}|$. The GVD parameter is also rescaled as,  $\delta_2=sgn(\beta_{20})+\chi\xi$, where, $\delta_2=\beta_2/|\beta_{20}|$ and $\chi=g\frac{L_D}{|\beta_{20}|}=g{t_0^2}/{|\beta_{20}|^2}$.

\subsection{Waveguide Description}

\begin{figure}[h!]
 \begin{center}
  \includegraphics[trim=0.38in 0.0in 0.7in 0.1in,clip=true,  width=92mm]{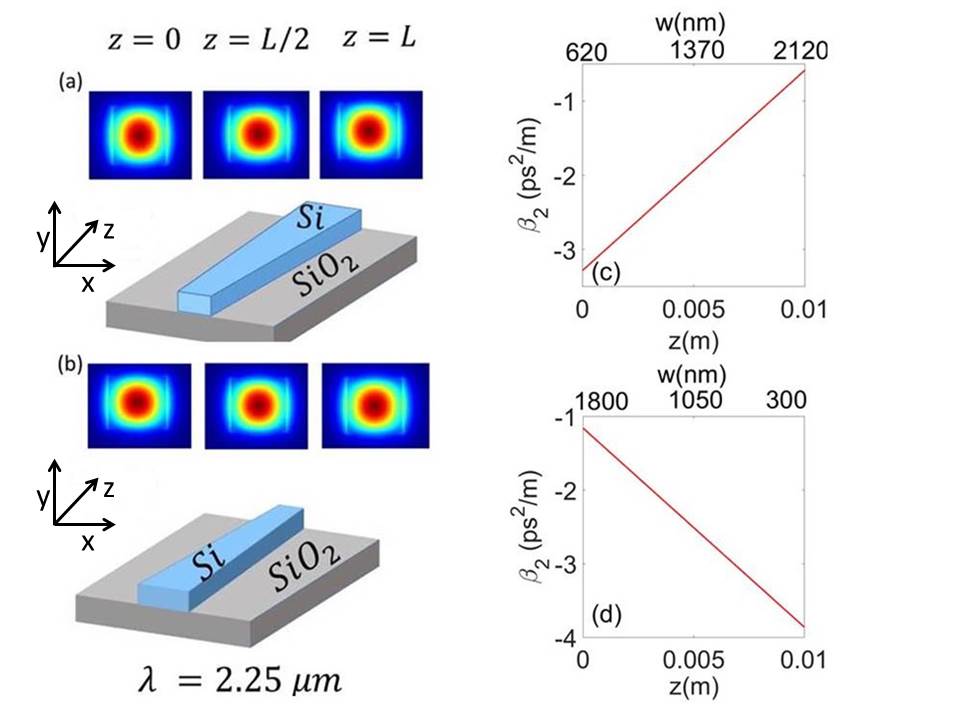}
  
  \vspace{-1em}
 \caption{ Schematic diagram of a Si-based slab waveguide with varying width ($w$).  (a) Width is increasing and (b) decreasing with distance.  The field distribution of the fundamental mode at $\lambda=2.25$ $\mu$m is also depicted at three different $z$ coordinate.  In plot (c) and (d) we demonstrate the linear variation of $\beta_2$ (which is calculated at $\lambda=2.25$ $\mu$m) with propagation distance $z$ and width $w$ for two waveguides. }
                 
\label{Figure1}
\end{center}
 \end{figure}

\noindent To investigate the dynamics of airy pulse under varying dispersion we consider the Si-based slab waveguide whose GVD profile can be tailored efficiently by manipulating the waveguide geometry. It is well known that the geometry of the slab waveguide is mainly controlled by two parameters, (i) slab height ($h$) and  (ii) slab width ($w$). One may achieve the desired GVD profile simply by manipulating $h$ and $w$. To obtain a linear spatial variation of $\beta_2 (z)$ for a fixed wavelength, we design a waveguide whose width $w$ varies linearly with the propagation length $z$ as $w=w_0+\epsilon z$, where $\epsilon$ denotes the rate of change of width with $z$ axis. In Fig.\ref{Figure1} we represent schematic diagrams of the waveguide of two distinct types, type-1 where the width is linearly increasing (plot a) and type-2 where the width is linearly decreasing (plot b) with propagation distance $z$. For type-1 waveguide, at  input the cross-sectional dimension is $w \times h$ = 620 nm $\times$ 800 nm and at output $w \times h$ = 2120 nm $\times$ 800 nm. For type-2 waveguide the dimensions are at input $w \times h$ = 1800 nm $\times$ 800 nm and at output $w \times h$ = 300 nm $\times$ 800 nm.    

   
  \noindent We consider $\epsilon = \pm 15\times10^{-5}$ which leads to a linear change in GVD as shown in Fig.\ref{Figure1} (c) and (d). For the proposed waveguides the rate of GVD change comes out to be $g \approx \pm 270 $ ps$^2$/m$^2$. The height ($h$) of the waveguide remains fixed at $800$ nm in all the cases. For type-1 and type-2 waveguide the dispersion length ($L_D$) becomes $\approx 2$ mm when we consider $t_0 = 90$ fs. We also consider the operating wavelength at $\lambda_0=2.25$ $\mu$m to avoid the detrimental two-photon absorption  (TPA) effect which is dominating for $\lambda <2.1 \mu$m in Si-based waveguide \cite{Bristow}.  In order to ensure that there is no nonlinear effect we compare the dispersion ($L_D$) and nonlinear length ($L_{NL}=1/\gamma_rP_0$) for the waveguides. The nonlinear parameter ($\gamma_r$) is defined as $\gamma_r=2\pi n_2/\lambda_0A_{eff}$, where $n_2$ is the Kerr coefficient. For silicon $n_2\approx3\times10^{-18} m^2W^{-1}$. The effective area of the confined mode is defined as, $A_{eff}=(\iint \limits_{-\infty}^{+\infty}|u(x,y)|^2dxdy)^2/\iint \limits_{-\infty}^{+\infty}|u(x,y)|^4dxdy$. For type-1 waveguide, $A_{eff}$  at the input and output are respectively, $\approx0.25 \mu$m$^2$ and $\approx$ 1.3 $\mu$m$^2$ which leads to $L_{NL}$ in the range of $0.30-1.55$ meters (for $P_0=100$ mW). Similarly for type-2 waveguide the $A_{eff}$ is calculated for the two ends are 
   $ \sim 1\mu$m$^2$ and $0.18 \mu$m$^2$ which leads to the range of $L_{NL}\approx 0.20-1.2$ meters (for $P_0=100$ mW).  Now for the proposed waveguide, $L_D\approx 2.0$ mm which leads to the condition $L_{NL}/L_D >>1$ throughout the waveguide length. The condition $L_{NL}/L_D >>1$ ensures that with the power level  $P_0=100 $ mW the proposed waveguides behave as a linear medium.

\subsection{Dynamics of FEAP and trajectory manipulation}

\begin{figure}[h!]
 \begin{center}
  \includegraphics[trim=0.0in 0.9in 0.0in 0.0in,clip=true,  width=90mm]{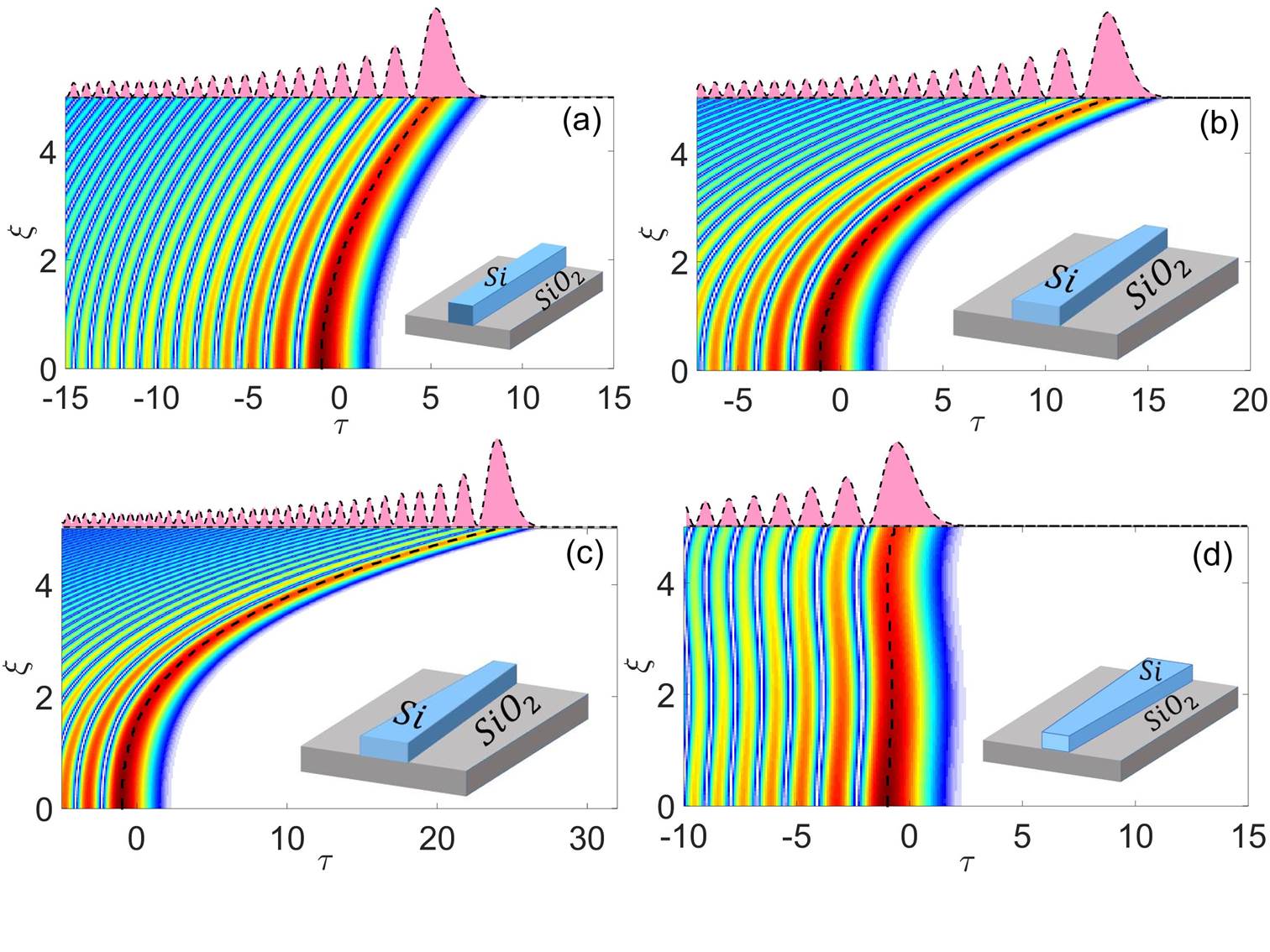}
  \includegraphics[trim=0.0in 2.4in 0.1in 1.2in,clip=true,  width=90mm]{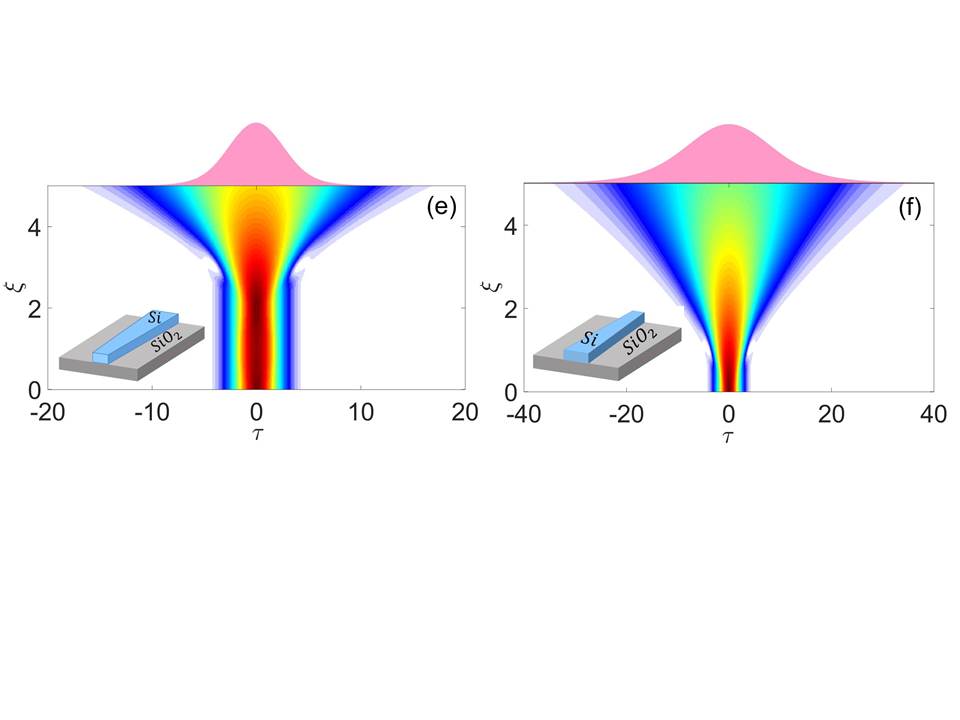}
  
  \vspace{-2em}
 \caption{Temporal density plots of the FEAP for different values of $\chi$, (a)$\chi=0$, (b)$\chi=-0.2$, (c) $\chi=-0.4$ and (d)$\chi=0.5$. On the upper panels of the figures we plot the analytical solution (black dashed lines) enclosing the numerical solution(pink shade). The analytical expression of the trajectory of the main lobe Eq. \ref{q3} is depicted (dashed line)on the mesh plot. We also compare  the trajectory of a sech pulse under linearly varying GVD for (e)$\chi=1$ and (f)$\chi=-1$.}
                 
\label{Figure2}
\end{center}
 \end{figure}

 We use a FEAP as input having a form $u(0,\tau)=Ai(\tau)\exp(a\tau)$, where $a$ is the truncation parameter that truncates the infinite energy pulse to a practically realizable finite energy pulse. The general solution of the governing equation (Eq.\ref{q1}) for a truncated airy pulse can be given as,  
 
\begin{equation}\label{q2}
u(\xi,\tau)=\exp(a^3/3-\widetilde{\alpha}\xi)Ai(b-n^2)\exp i\left(\frac{2}{3}n^3-nb\right) 
\end{equation}
where, $b=(\tau-a^2)$ and $n=ia-\frac{\xi}{2}+\chi\frac{\xi^2}{4}$. The dynamics of the FEAP  is illustrated in Fig.\ref{Figure2} where we demonstrate the  density distribution of the propagating pules for different values of GVD rate $\chi$.  As illustrated in the density plots, the parameter  $\chi$ significantly influences the trajectory and final temporal position of the propagating airy pulse. The airy pulse does not follow the usual ballistic trajectory when the width of the waveguide is decreasing or increasing with distance which accounts for a non-zero $\chi$ (Fig.\ref{Figure2}(b)-(d)). The temporal position of the main lobe ($\tau_p$) evolves as,
 
\begin{equation}\label{q3}
  \tau_p (\xi)=\tau_{0p}+\frac{ \xi^2}{4}\left(\frac{\chi\xi}{2}-1\right)^2
  \end{equation}
where $\tau_{0p}\approx -(3 \pi/8)^{2/3}$ is the initial temporal position of the primary lobe of the pulse. Eq.(\ref{q3}) provides the theoretical estimation of the trajectory of the main lobe of FEAP.  In Fig.\ref{Figure2} we demonstrate the overall dynamics of Airy pulse which we obtain by solving the Eq.(\ref{q1}) numerically using split-step Fourier method \cite{Agarwal}. From the figures it is evident that the dynamics of FEAP is affected significantly by the GVD rate $\chi$. The usual ballistic trajectory of the Airy pulse  deforms under varying GVD. The trajectory of the main lobe can be controlled by the GVD rate $\chi$. Note, $\chi$ can be positive or negative and by changing its numeric value one can manipulate the trajectory. The airy pulse decelerates more when the decreasing rate of waveguide width is large (see plot (b) and (c)). It is obvious from Eq. \eqref{q3}, that airy pulse will always decelerate for $\chi<0$. However the usual parabolic airy dynamics is almost lost when $\chi>0$ and we observe a quasi-linear trajectory (see plot (d)). It is interesting to note that, for a waveguide of length  $L$ the main lobe retain its position at output for $\chi=2/L$ . We superimpose the analytically obtained trajectory of the main lobe (black dashed lines) based on the Eq.(\ref{q3}) in numerical mesh plots and obtain a perfect agreement. In the top panel of the Fig. \ref{Figure2} we depict the shape of the airy pulse at output which we obtain numerically (shaded curve). The analytical solution of the propagating truncated airy pulse (see Eq. (\ref{q2})envelopes the shaded curve through dashed line. We also compare the dynamics of a $sech$ pulse for varying dispersion in Fig.\ref{Figure2}(e) and (f). The variation of the GVD parameter does not affect the trajectory of $sech$ pulse that only experiences a temporal broadening.

We know that the energy distribution of FEAP does not remain intact while propagating inside an optical medium as this is not the natural solution of the dispersive system. A constant decay of the peak power of a FEAP is incurred through the truncation parameter $a$. FEAP also experiences a linear material loss ($\alpha$) which is typically $\sim$ 0.6 dB/cm  for Si \cite{zou,Mashanovich}. In application point of view, it is desired that the airy pulse should retain its shape and power level at the output. We observe that the rate of energy loss of a propagating FEAP can also be manipulated through dispersion engineering.

   
 \begin{figure}[h!]
 \begin{center}
  \includegraphics[trim=0.0in 1.4in 0.0in 1.2in,clip=true,  width=88mm]{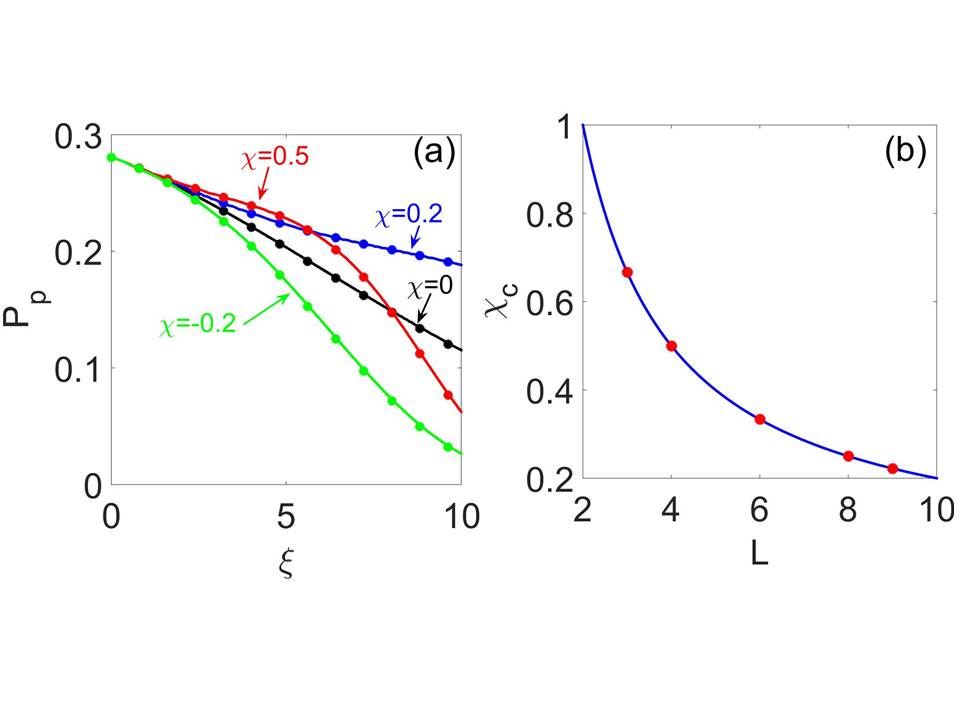}
  
  \vspace{-1em}
 \caption{(a) The variation of the peak power of the FEAP ($P_{p}$) with $\eta$  for different values of $\chi$. The solid lines represent the analytical form where as the dots are numerical data points. It can be seen that for $\chi=0.2$ the attenuation of  $P_{p}$ is minimum. (b) The variation of $\chi_c$ for waveguides of different length $L$. Blue solid line represents the analytical results where as red solid dots shows the corresponding numerical results.}
                 
\label{Figure3}
\end{center}
 \end{figure}

In Fig.\ref{Figure3}(a) we demonstrate the rate of the attenuation of the peak power ($P_{p}$) of the main lobe of FEAP with distance considering linear loss as well. It is observed that the rate of attenuation is different for non-identical $\chi$ values. We can see that for a particular value of $\chi$ the peak power ($P_{p}$) reduces less  ($\chi=0.2$ in Fig.\ref{Figure3}(a)) while for all the other cases the power attenuates at a relatively higher rate.  The variation of $P_{p}$ against propagation distance $\xi$ can be written in the form, $P_{p}(\chi,\xi)=P_{p}^{(0)}e^{-\Sigma(\xi)}$, where $P_{p}^{(0)}$ is the peak power at input and  $\Sigma(\xi)=2\widetilde{\alpha}\xi+a\frac{\xi^2}{2}(\frac{\chi \xi}{2}-1)^2-\frac{2}{3}a^3$. From the expression it is obvious that, the peak power attenuates monotonically due to the presence of the material loss. However, the overall attenuation can be engineered through $\chi$. Minimizing the decay factor $\Sigma$ for $\chi$ we obtain a optimized relation $\chi_c=\frac{2}{L}$ for which we expect minimal power decay given waveguide length ($L$) we can always find a critical GVD rate $\chi_c$ for which the power decay is minimal. In Fig.\ref{Figure3}(a) we illustrate the variation of $P_{p}$ for a waveguide of length $L=10$ and obtain the minimal power decay for $\chi_c=0.2$ which is consistent with the theoretical prediction.  We extend our simulation for different waveguide length and numerically obtain  the corresponding $\chi_c$ (dots) for which maximum power transform occurs. The numerical result (dots)                corroborate well with the theoretical expression (solid line) which we obtain by minimizing $\Sigma$ as shown in Fig.\ref{Figure3}(b).

\section{Dynamics of FEAP under Periodic GVD}

We demonstrate that, the geometry of the waveguide affects the dynamics of FEAP through dispersion. One can think of the geometry of the waveguide as a effective tool to manipulate the airy dynamics. We extend this idea in this section where we investigate the propagation properties of a FEAP inside the waveguide with periodic width variation. A periodically varying width of the waveguide leads to oscillating GVD over distance \cite{Bai}. We study the airy dynamics for oscillating GVD as well as under TOD where pulse experiences periodic singularity.

\subsection{Waveguide description}

\begin{figure}[h!]
 \begin{center}
  \includegraphics[trim=0.8in 0.00in 1.2in 0.2in,clip=true,  width=80mm]{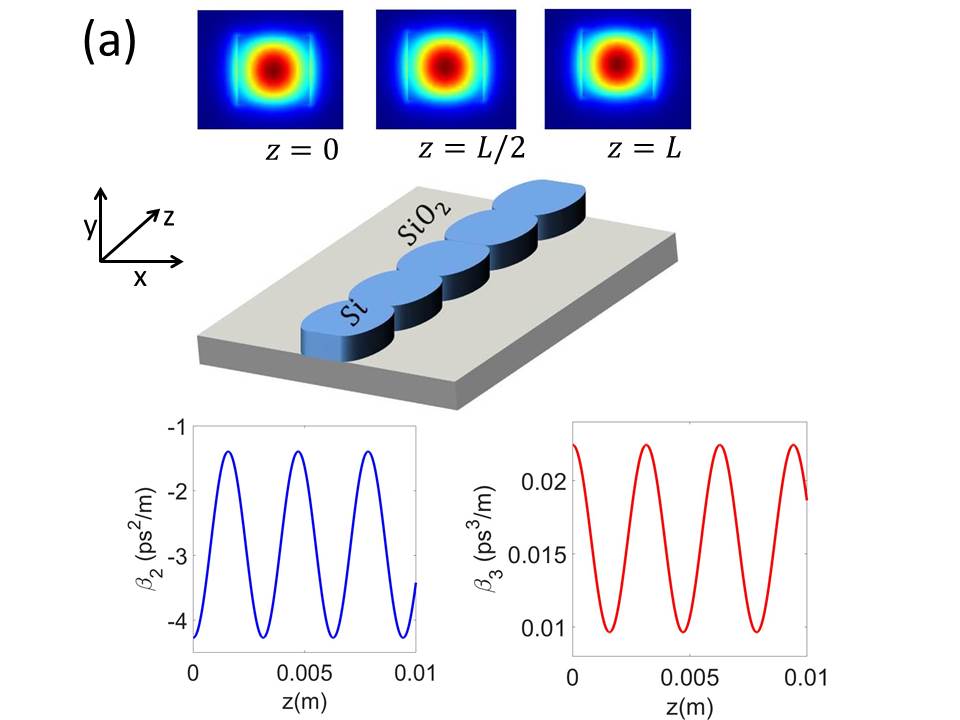}
  \includegraphics[trim=0.8in 0.00in 0.8in 0.2in,clip=true,  width=80mm]{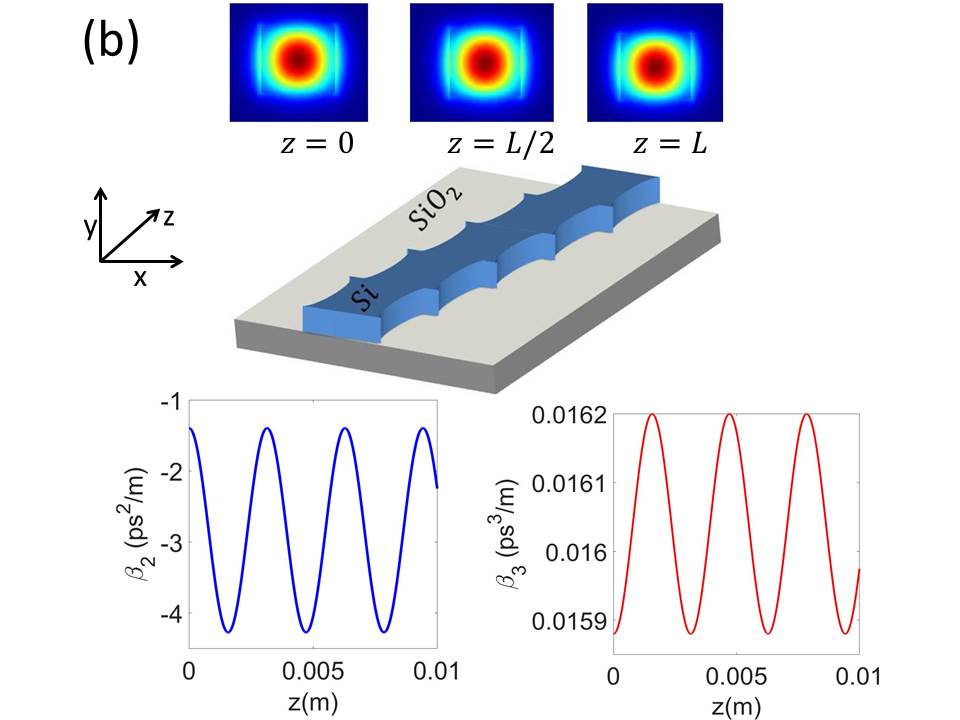}
  
  \vspace{-1em}
 \caption{ Schematic diagram of a Si-based slab waveguide with oscillating width ($w$).  (a) Width is varying periodically with distance starting from a lower value to a higher value, (b) the width decreases first and then increases with distance.  The field distribution of the fundamental mode at $\lambda=2.25$ $\mu$m is also depicted at three different $z$ coordinate.  In the plots we demonstrate the periodic variation of $\beta_2$ and $\beta_3$ (which is calculated at $\lambda=2.25$ $\mu$m) with propagation distance $z$. }
                 
\label{Figure4}
\end{center}
 \end{figure}

\noindent  A periodic width ($w$) variation of a waveguide leads to oscillating GVD profile \cite{Bai}. We design the waveguide considering the width variation as, $w=w_0+\epsilon \cos(z/z_0)$, where $w_0=870$ nm and the value of the strength $\epsilon=800$ nm.  The period of the oscillation is $z_0$= 500  $\mu$m. Depending on the numeric value of the $\epsilon$, we can consider two types of waveguide. In Fig.\ref{Figure4},type-1(plot a) where $\epsilon$ is positive and the width  increases at first and then decreases and type-2(plot-b) where the opposite effect occurs with negative $\epsilon$. For the waveguide design we keep the height of the waveguide fixed at $800$nm. For type-1 and type-2 waveguides we calculate the GVD and TOD profiles for fundamental modes using commercial COMSOL software which exhibit sinusoidal variation as shown in Fig.\ref{Figure4}.     

\subsection{Dynamics of FEAP under periodic GVD}    

\begin{figure}[h!]
 \begin{center}
  \includegraphics[trim=2.0in 0.0in 1.6in 0.0in,clip=true,  width=82mm]{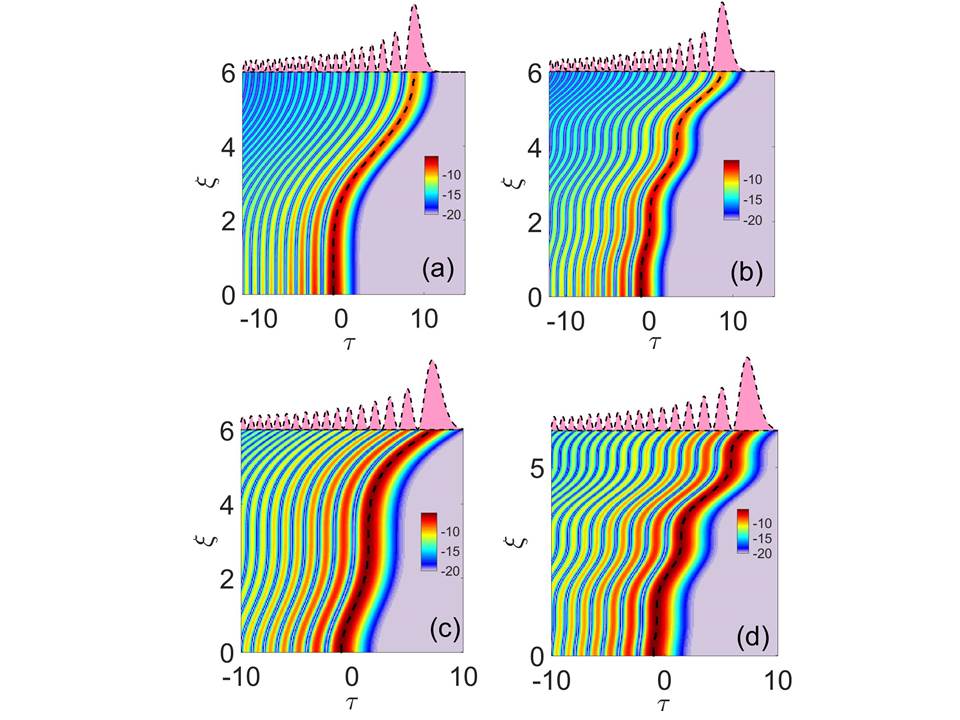}
  \includegraphics[trim=1.8in 2.5in 1.6in 1.2in,clip=true,  width=82mm]{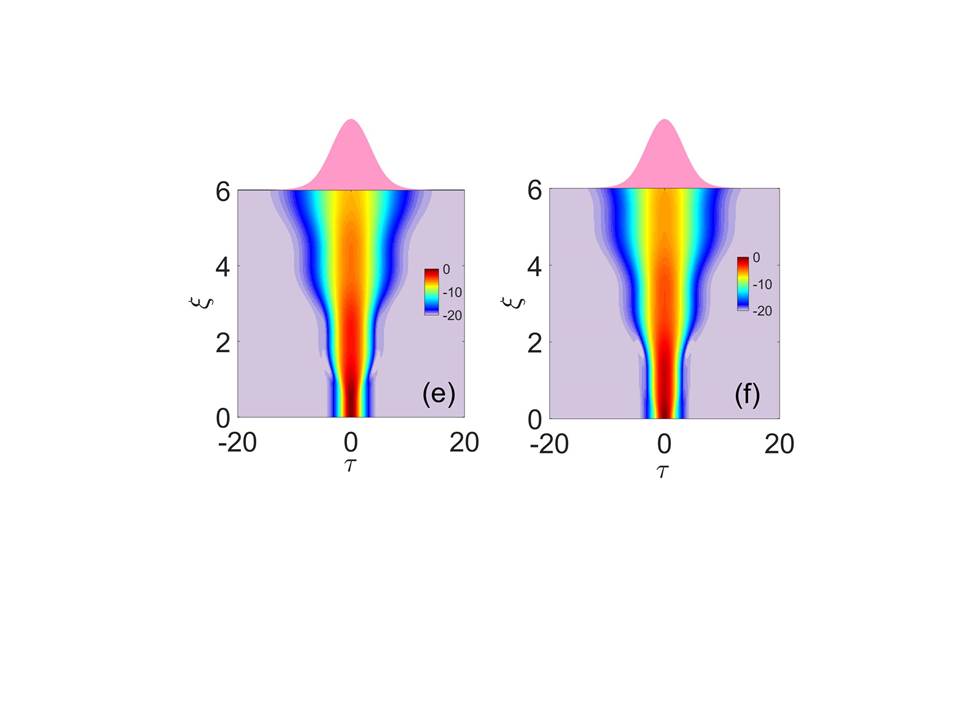}
  \vspace{-1em}
 \caption{Temporal density plots of the propagating FEAP for different values of $\chi$ and $\mu$. For the upper row $\chi=1$ and for(a) $\mu=1$ and for (b)$\mu=3$. In the lower row $\chi=-1$ and  for (c)$\mu=1$ and for (d) $\mu=3$. In the upper panels we show the analytical solutions (black dashed line) which enclose the the numerical output (pink shaded area). The trajectory of the primary lobe is plotted (black dashed lines)in each density plots that we find analytically. Finally we plot the dynamics of sech pulse for (e)$\chi=1$ and (f)$\chi=-1$ with $\mu=3$. }
                 
\label{Figure5}
\end{center}
 \end{figure}

In the previous section we investigate the dynamics of a FEAP for a waveguide having  dispersion that varies linearly with propagation direction $z$.  Now if the width of the waveguide varies periodically with its length then as a consequence we have the periodic GVD parameter \cite{Bai}. In such case, the form of $\beta_2$ can be written as $\beta_2(z)=\bar{\beta}_{20}+f \cos(\bar{\mu}z)$, where the parameters $f$ and $\bar{\mu}$ account for the strength and frequency of periodicity respectively. $\bar{\beta}_{20}$ is the average value of GVD.  The GVD parameter can be rescaled in normalised unit as, $\delta_2(\xi)=sgn(\bar{\beta}_{20})+\chi \cos(\mu \xi)$, where $\chi=f/|\bar{\beta}_{20}|$ ,  $\mu=\bar{\mu}L_D$ and $L_D=t_0^2/|\bar{\beta}_{20}| $. Taking the normalised form of GVD parameter if we solve the governing equation (Eq.\ref{q1}) the solution comes out to be a similar looking form that we obtain in (Eq.\ref{q2}),  
\begin{equation}\label{q4}
u(\xi,\tau)=\exp(a^3/3-\widetilde{\alpha}\xi)Ai(b-m^2)\exp i\left(\frac{2}{3}m^3-mb\right) 
\end{equation}
where, $m=ia-\frac{\xi}{2}+\frac{\chi}{2\mu}\sin(\mu\xi)$.  The propagating pulse now has more degrees of freedom and the trajectory of the pulse can be manipulated by changing the amplitude ($\chi$) and period ($\mu$) of the GVD parameter. In Fig.\ref{Figure5} we  demonstrate the  density plots of the propagating FEAP under different set of $\chi$ and $\mu$. The trajectory of the FEAP in  presence of oscillating GVD can be derived as, 
\begin{equation}\label{q5}
  \tau_p=\tau_{0p}+\frac{\xi^2}{4}+\frac{\chi}{2\mu}\sin(\mu\xi)\left(sgn(\bar{\beta}_{20})\xi+\frac{\chi}{2\mu} \sin(\mu\xi)\right).
  \end{equation} 
Due to the periodic variation of GVD, the temporal position of the Airy pulse is now  oscillating. In Fig.\ref{Figure5} we superimpose the trajectory that is obtained theoretically (black dashed lines) using Eq. \eqref{q5} which agrees well with numerical results.The periodic variation of the GVD parameter does not affect the trajectory of $sech$ pulse that only experiences a temporal broadening as it moves through the waveguide(Fig.\ref{Figure5}(e)-(f)). 

   
 \begin{figure}[h!]
 \begin{center}
  \includegraphics[trim=0.65in 0.0in 1.0in 0.0in,clip=true,  width=88mm]{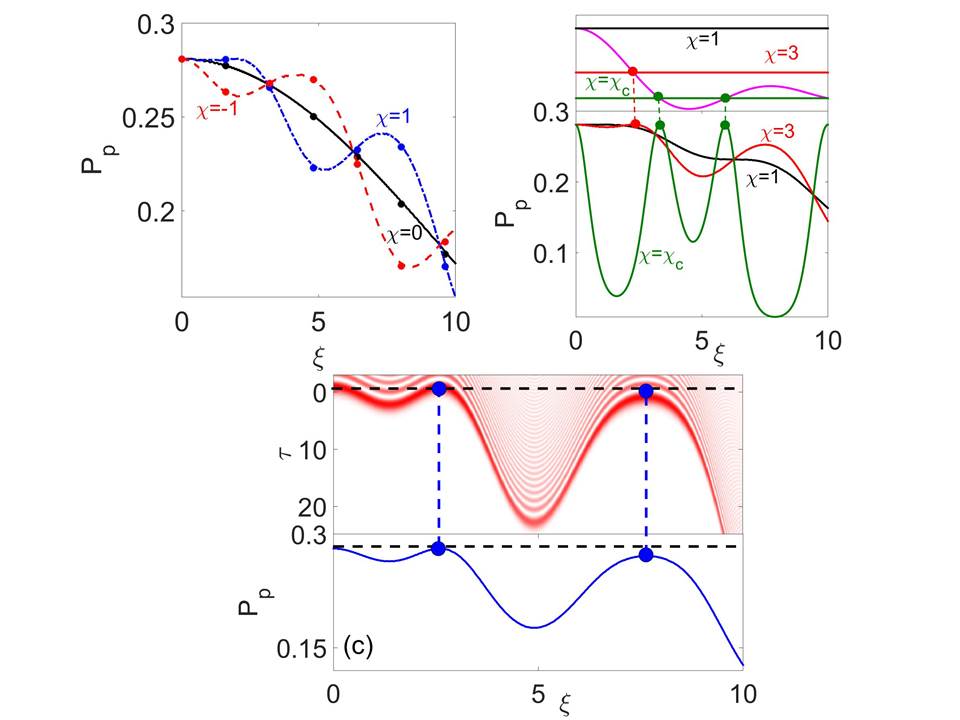}
  
  \vspace{-1em}
 \caption{(a) The variation of the peak power of FEAP with the propagating distance for  different strength of oscillating GVD parameter $\chi$. The lines represent the analytical expression where as dots are the corresponding numerical data. (b) The variation of $P_{p}$  for different $\chi$ with fixed $\mu=1$. In the upper panel we plot the functions constituting  the transcendental equation whose solution indicates the value of $P_p$ which is identical to its initial value. The solutions are indicated by the dots for different $\chi$.  (c) The figure indicates the relationship between the temporal position of the main lobe of the Airy pulse with the oscillating power. It can be seen that the power reaches to its value exactly at the same $\xi$ when pulse returns to its initial temporal location.   }
                 
\label{Figure6}
\end{center}
 \end{figure}

The numerical solution also reveals that, in absence of loss peak power $P_{p}$ varies periodically when GVD is oscillating. For a lossless truncated airy pulse, we derive the expression of the $P_{p}$ as, 
 \begin{equation}\label{q6}
 P_{p}(\xi,\mu,\chi)=P_{p}^{(0)} e^{- \Gamma^2},
 \end{equation}
 where the decay factor $\Gamma$ is given as, $\Gamma=\sqrt{\frac{a}{2}}\int\limits_{0}^{\xi}\delta_2(\xi)d\xi.$ Using the explicit form of $\delta_2$ one can quantify the decay factor as, $\Gamma=\sqrt{\frac{a}{2}}[sgn(\bar{\beta}  _{20})\xi +\frac{\chi}{\mu}\sin(\mu\xi)]$. Since the decay factor is periodic we can expect an oscillatory evolution of the peak power of the main lobe of the propagating FEAP. The periodic nature of the decay factor $\Gamma$ leads to more interesting features. For example, at $\xi=n \pi/\mu, (n=1,2,3,4..)$   the peak power will be identical irrespective of GVD profile. We illustrate this feature in Fig.\ref{Figure6}(a). It is interesting to note that the decay factor $\Gamma$ can vanish for a specific propagation length ($\xi_c$) satisfying the transcendental equation $\sin(\mu \xi)/\mu \xi=-sgn(\beta_{20})\chi$. At $\xi_c$ the $\Gamma$  is zero and peak power retains to its input value. In Fig.\ref{Figure6}(b) we plot the variation of $P_{p}$ with distance for different modulation strength $\chi$. A special value of $\chi$ can be chosen such a way that the initial power is revived at the output which is shown in Fig.\ref{Figure6}(b). In absence of amplitude modulation of GVD (i.e $\chi=0$), the peak power decays monotonically. For $\chi\neq 0$, the variation of $P_{p}$ becomes oscillatory. It is also illustrated that how for a particular value of $\chi=\chi_c$ the peak power carried by the primary lobe of FEAP revives to its original value at a fixed propagation length. This specific length can be unique or many valued depending on the single or multi-valued  solution of the transcendental equation. The oscillatory dispersion  profile affects the peak power and temporal location of the main lobe of FEAP in a complimentary manner. In Fig.\ref{Figure6}(c)  using density plot we demonstrate the variation of the temporal location ($\tau_p (\xi)$) of the main lobe with propagation distance $\xi$.  The temporal position should follow the path as derived in Eq. \eqref{q5} which suggests the usual balletic propagation of FEAP is no longer valid under modulated GVD profile. The FEAP oscillates against its initial position. The modulation strength $\chi$ is kept to a value $\chi=5$. In the same plot we demonstrate the variation of the $P_{p}$ which is oscillating over distance. It is interesting to note that the oscillating peak power and temporal position both revives to its initial values exactly at the same space point. This is an important piece of information in the context of application.

\subsection{Selective focusing under periodic TOD}    

\begin{figure}[h!]
 \begin{center}
  \includegraphics[trim=1.2in 0.2in 2.0in 0.1in,clip=true,  width=86mm]{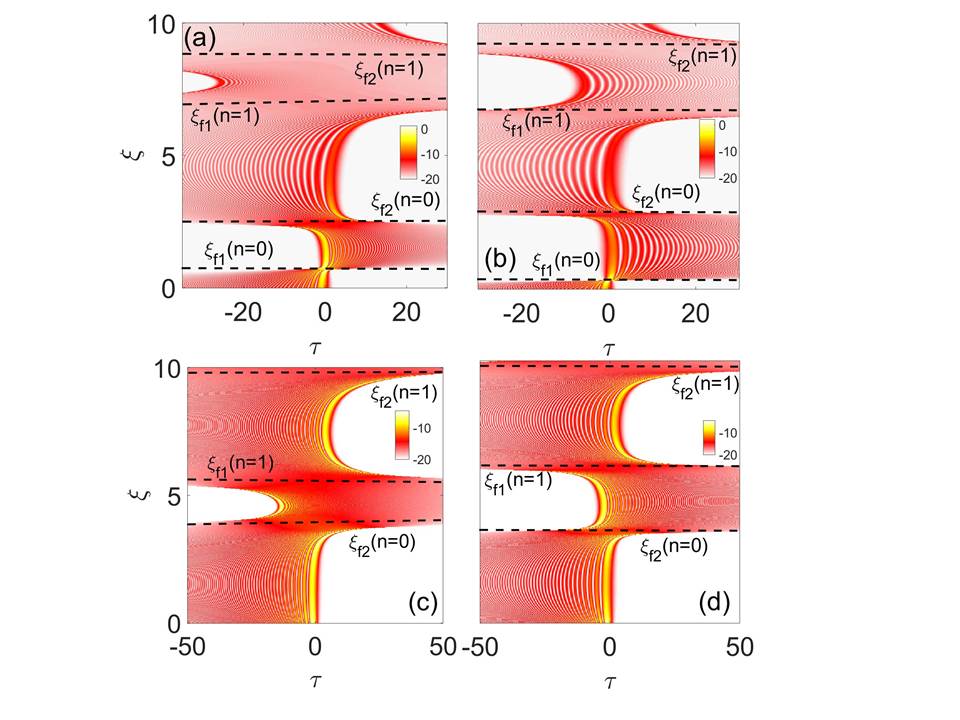}
  
  \vspace{-1em}
 \caption{The density plot of FEAP in presence of periodically varying TOD parameter with different strength (a)$\chi_3=0.5$ (b)$\chi_3=1$ with $\chi=-0.5$ which is for type-2 oscillating waveguide. For type-1  waveguide the density plots are shown for (c)$\chi_3=-0.5$ (d)$\chi_3=-1$ with $\chi=0.5$. The phenomenon of multiple flipping can be seen from the figure and the positions of flipping are indicated by the dashed lines which is obtained from Eq.\eqref{q9}.  }
                 
\label{Figure7}
\end{center}
 \end{figure}

In the previous sections we  ignore the effects of the higher order dispersions during the study of the dynamics of FEAP under modulated GVD profile. However, if the pulses are launched near zero GVD wavelength then the effect of TOD will be significant. The dynamics of FEAP in presence of moderate and strong TOD has been a topic of investigation lately\cite{driben}. It has been shown that FEAP shows peculiar behaviour in presence of TOD. For positive TOD  coefficient ($\delta_3>0$) the FEAP focuses to a gaussian pulse after moving a specific distance and the temporal distribution flips \cite{driben}. The position and the area of this focusing zone mainly depends on the numeric value of $\delta_3$. Some works have also been done where it is demonstrated that this flipping phenomenon can also be controlled by external parameters (like phase modulation) which are independent of the TOD coefficient \cite{Roy_b}. In our work we show that the geometry of the waveguide plays a pivotal role in the dynamics. The periodic variation of the waveguide geometry leads to a periodic variation of  the TOD coefficient $\beta_3$.  In real unit $\beta_3$ can be expressed as $\beta_3(z)=\bar{\beta}_{30}+q\cos(\bar{\mu}z)$. $\bar{\beta}_{30}$  and $\bar{\mu}$ represent the average value of TOD parameter and period of the oscillation, respectively. The strength of the modulation is controlled by the factor $q $. Including TOD term the governing equation can be written as 
\begin{equation} \label{q7}
i\frac{\partial u}{\partial \xi}=\frac{{{\delta}_{2}(\xi)}}{2}\frac{{{\partial }^{2}}u}{\partial {{\tau}^{2}}}+i\delta_{3}(\xi)\frac{{{\partial }^{3}}u}{\partial {{\tau}^{3}}}-i\widetilde{\alpha}\xi
\end{equation}
where, $\delta_3(\xi)=\bar{\delta}_{30}+\chi_3 \cos (\mu\xi)$ is the distance dependent TOD parameter in normalised unit.  The strength of the modulation ($\chi_3$) in normalised unit can be written as $\chi_3=\frac{q}{6t_0|\bar{\beta}_{20}|}$, where $\bar{\delta}_{30}=\frac{\bar{\beta}_{30}}{6t_0|\bar{\beta}_{20}|}$. The period $\mu$ is rescaled as $\mu=\bar{\mu}L_D$.  The general solution  of Eq. \eqref{q7},

\begin{equation}\label{q8}
\begin{aligned}
 u_a(\xi,\tau)=\frac{1}{c}\exp\left(a^3/3-\widetilde{\alpha}\xi\right)Ai\left(\frac{b}{c}-\frac{m^2}{c^4}\right)\\
 \exp i\left(\frac{2m^3}{3c^6}-\frac{mb}{c^3} \right),
     \end{aligned} 
     \end{equation}  
 where $c=(1-3\bar{\delta}_{30}\xi-3\frac{\chi_3}{\mu}\sin\mu\xi)^\frac{1}{3}$.  From the solution it is evident that  a singularity appears at $c=0$ which leads to a transcendental equation $\sin(\mu\xi)=\frac{\mu}{3 \chi_3}(1-3\bar{\delta}_{30} \xi)$.  For a specific case when $\bar{\delta}_{30}=0$ the singularity condition simplifies to   $\sin(\mu\xi)=\frac{\mu}{3\chi_3}$.  At singular point the original  FEAP reshapes to form a Gaussian pulse and flips temporally. The singularity condition ($c=0$) gives rise to two sets of flipping positions,     
\begin{equation}\label{q9}
\begin{aligned} 
\xi_{fj}^{(n)} = (-1)^{j-1}\frac{1}{\mu}\sin^{-1}\left(\frac{\mu}{3\chi_3}\right)+\frac{\pi}{\mu} [2n+(j-1)] \ \ (j=1,2  ) 
\end{aligned}
\end{equation}
with $n=0,1,2,.....$ , where FEAP loses its identity. From the expression of Eq. \eqref{q9} it is clear that the FEAP will face multiple flipping while moving in a medium with periodic TOD. In Fig. (\ref{Figure7})  we demonstrate the evolution  of a FEAP under periodic TOD. It is evident that the pulse experiences multiple temporal flippings at the specific locations  estimated theoretically by  Eq.\eqref{q9}. The input FEAP first faces a singularity at $\xi_{f1}^{(0)}$ for which $c=0$ in Eq.(\ref{q8}). At this specific point the FEAP turns into a Gaussian pulse and after that it propagates as a FEAP with temporally flipped wings upto the next flipping point $\xi_{f2}^{(0)}$.  This phenomenon repeats itself as the pulse moves forward. To get more insight about this peculiar dynamics of the FEAP under periodic TOD, we try to find the analytical solution of the pulse at different zones of propagation. The solution beyond the flipping point for static TOD  is already reported \cite{Roy}. We exploit this concept to find a general solution for periodically varying TOD. Careful investigation reveals that the flipping areas corresponding to different values of $n$  in Fig.\ref{Figure7} are not of same size. In fact the area of flipping region increases when $n$ increases.  We derive the general form of the Gaussian pulse at flipping points as,
 
   
 \begin{figure}[h!]
 \begin{center}
  \includegraphics[trim=0.2in 0.8in 0.5in 1.1in,clip=true,  width=80mm]{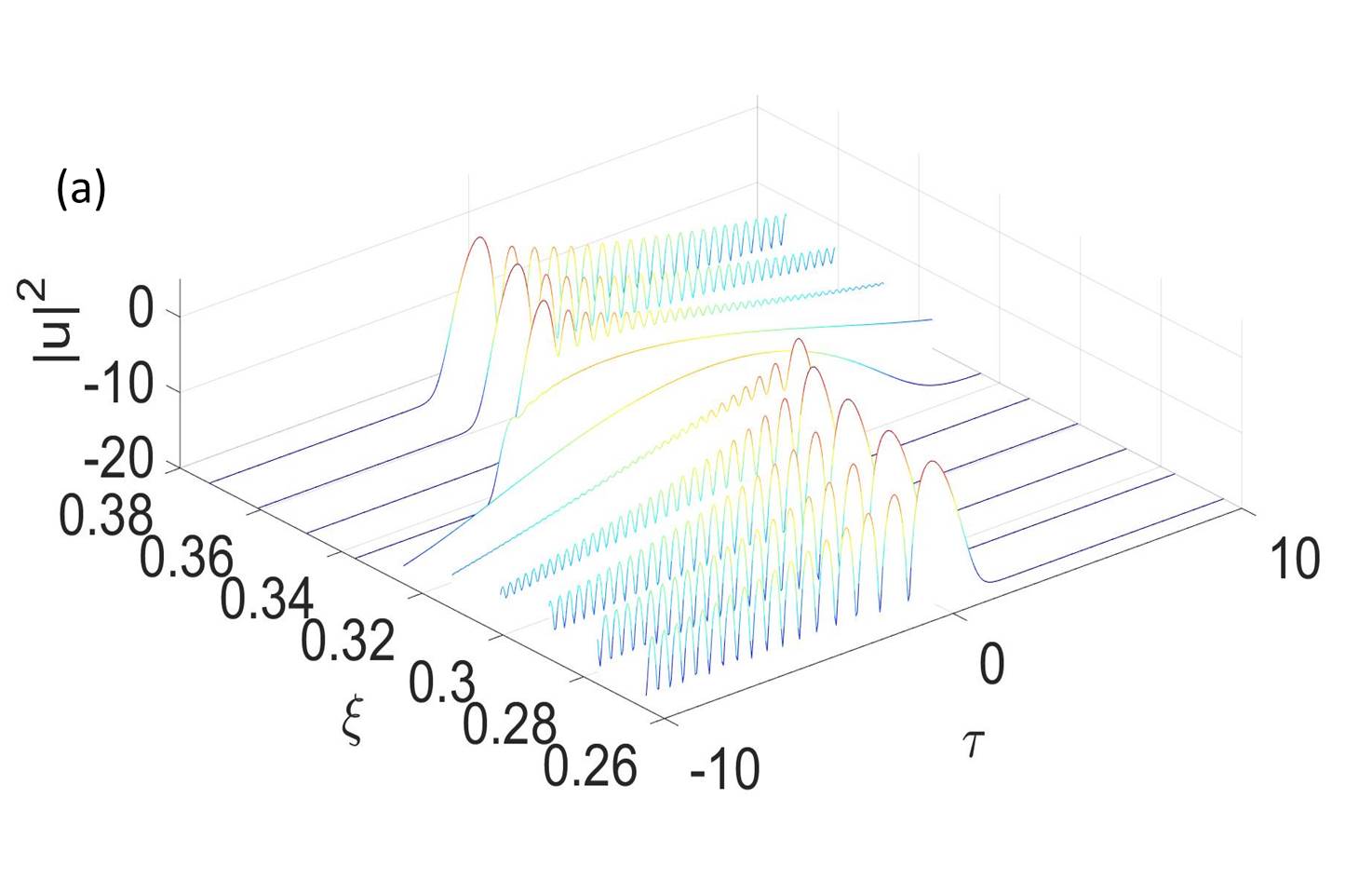}
  \includegraphics[trim=0.1in 0.0in 0.4in 0.0in,clip=true,  width=80mm]{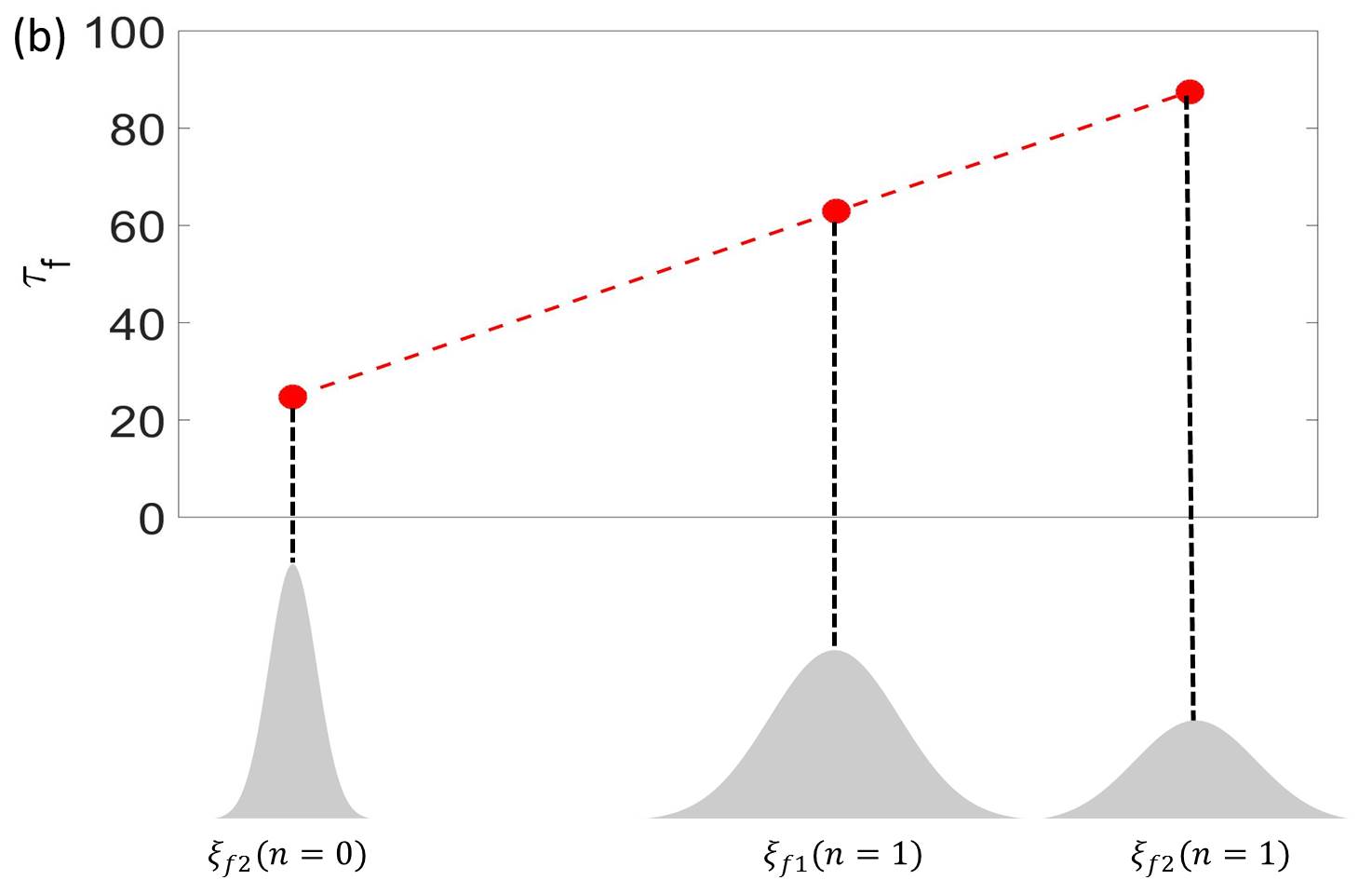}
  \includegraphics[trim=0.0in 2.0in 0.3in 0.9in,clip=true,  width=80mm]{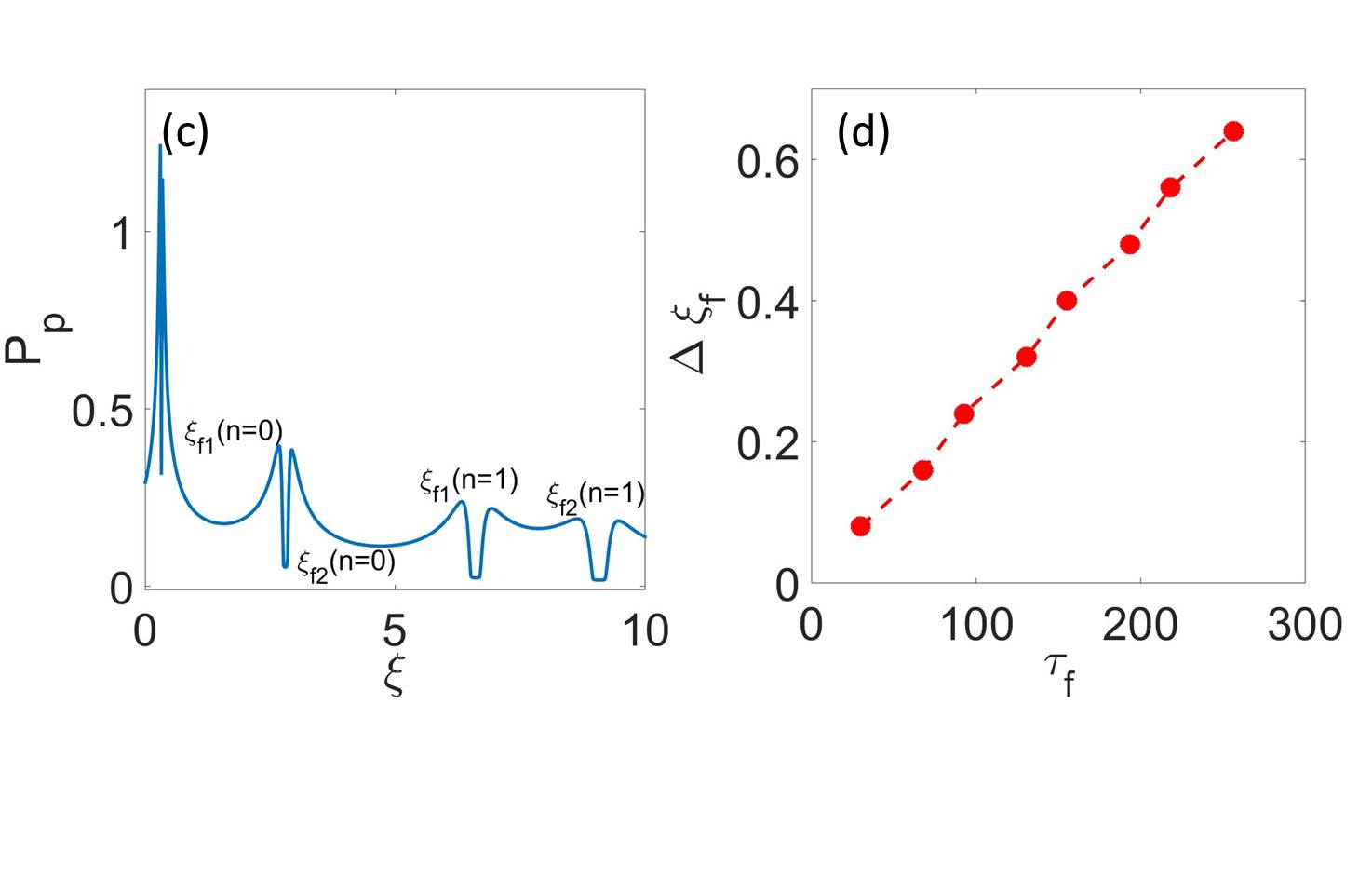}
  \vspace{-1em}
 \caption{(a) The propagation of the FEAP at flipping zone is highlighted where the pulse converges to a Gaussian pulse at the flipping position and then it propagates again with inverted temporal wings.(b) The variation of the width $\tau_f$ of the Gaussian pulses obtained at different flipping positions. The shape of the Gaussian pulses are also given at the bottom of the plot. The width enhances as we go to the higher order flipping positions.  (c) The variation of $P_p$ with $\xi$ for $\chi_3=1$. The dipping of $P_0$ indicates the position of flipping. It can be seen that the length of flipping area enhances for higher order flipping positions.(d) The variation the length of flipping area $\Delta \xi_f$ with $\tau_f$ of the Gaussian pulse obtained at the flipping positions. It can be seen that higher $\tau_{f}$ enhances $\Delta \xi_f$.  }
                 
\label{Figure8}
\end{center}
 \end{figure}

   
 \begin{figure}[h!]
 \begin{center}
  \includegraphics[trim=1.2in 0.2in 2.0in 0.1in,clip=true,  width=80mm]{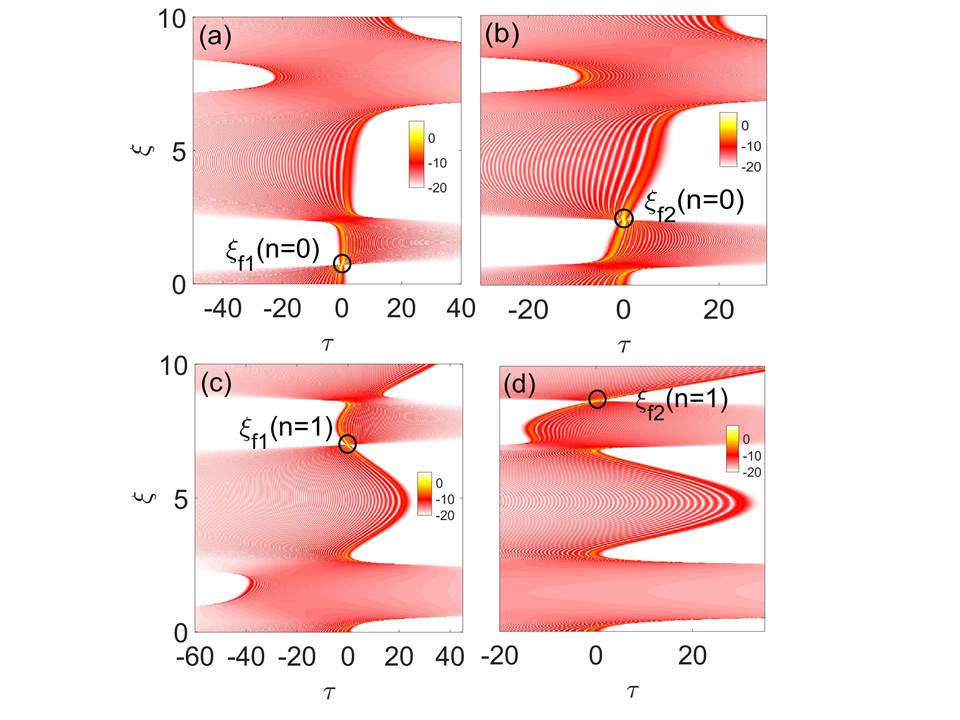}
  
  \vspace{-1em}
 \caption{Selective focusing for $\chi_3=0.5$. The black circles represent the tight focusing positions. The focusing takes place when the condition of Eq.\ref{q12} is achieved. We can select the position of the tight focusing by suitably adjusting the parameter.}
                 
\label{Figure9}
\end{center}
 \end{figure}

 \begin{equation}\label{q10}
 u_b(\xi_f,\tau)=U_0 \exp\left[-\frac{(\tau-a^2)^2}{\tau_f^2}\right]\exp(i\phi),
 \end{equation}
    where $U_0= \frac{1}{2\sqrt{\pi\gamma}}\exp(a^3/3)$ and $\phi=\frac{1}{2}\tan^{-1}\left(\frac{\Delta^{(n)}}{a}\right)-\frac{\Delta^{(n)}(\tau-a^2)^2}{4\gamma^2}$. The parameter $\Delta^{(n)}$ is defined as $\Delta^{(n)}=\frac{1}{6}\left[\frac{\chi}{\chi_3}-3\xi_{fj}^{(n)}\right]$ and $\gamma=\sqrt{a^2+\Delta^{(n)2}}$. The characteristic width of the Gaussian pulse is $\tau_f=2\gamma/\sqrt{a}$. It is evident that the width of the Gaussian pulse will  differ at different flipping positions depending on the values of $\xi_{fj}^{(n)}$ which can be found from Eq(\ref{q9}). In Fig.\ref{Figure8}(a) we plot the dynamics of FEAP around the first flipping position $\xi_{f1}^{(0)}$ where it can be clearly observed that the pulse merges to gaussian pulse before its temporal wings flip. It is interesting to note that, the length of the flipping zone  depends on the width of the Gaussian pulse generated in the flipping point\cite{Roy}. Greater the width greater is the length of the flipping area. For the Gaussian pulse the  full width at half maxima $\tau_{FWHM}$ is given as,
    
\begin{equation}\label{q11}
\tau_{FWHM}=2\sqrt{2 \ln2}\left(a+\frac{\Delta^{(n)2}}{a}\right)^\frac{1}{2}
\end{equation}   
The expression suggests that the width of the pulse depends on the position of the flipping and it increases as we go to the higher orders  of $n$.  In(Fig.\ref{Figure8}(b)) we plot the variation of the temporal width $\tau_f$ at different flipping points where individual Gaussian pulses are emerged. It is evident from the illustration that,the widths of the Gaussian pulse gradually increases at each flipping point denoted by $\xi_{fj}^{(n)}$. The  evolution of $P_p$ for $\chi_3=1$ is plotted in Fig.\ref{Figure8}(c) where we can see that the maximum peak-power carried by the pulse dips down at the position of flipping. The length of the valley shown in the Fig.\ref{Figure8}(c) measures the flipping length $\Delta \xi_{fj}^{(n)}$ which enhances at each flipping point.  The flipping length $\Delta \xi_{fj}^{(n)}$ is proportional to the width of the Gaussian pulse generated at $\xi_{fj}^{(n)}$. In \ref{Figure8}(d) we numerically demonstrate the relationship between the Gaussian width $\tau_f$ and flipping length $\Delta \xi_f^{(n)}$ which is almost linear.

\noindent Note, we can approximate the expression of gaussian width as $\tau_{FWHM}\approx \Delta^{(n)}\sqrt{2\ln2/a} $ for small truncation parameter $a$.  Now, for $\Delta^{(n)}$ the width of the Gaussian pulse obtained at the flipping point nearly vanishes which is the condition for absolute temporal focusing. In the neighbourhood of the absolute focusing point the peak power of the propagating FEAP reaches to its maxima which may be useful in application point of view.  The uniqueness of the oscillating GVD parameters lies in the fact that here we can selectively focus the FEAP to a point by varying the amplitude ($\chi$) of periodic GVD parameter. The parameter $\chi$ for which $\Delta^{(n)}=0$ is 

\begin{equation}\label{q12}
\chi=3\chi_3\xi^{(n)}_{fj}
\end{equation}

   
 \begin{figure}[h!]
 \begin{center}
  \includegraphics[trim=2.0in 0.2in 1.7in 0.0in,clip=true,  width=82mm]{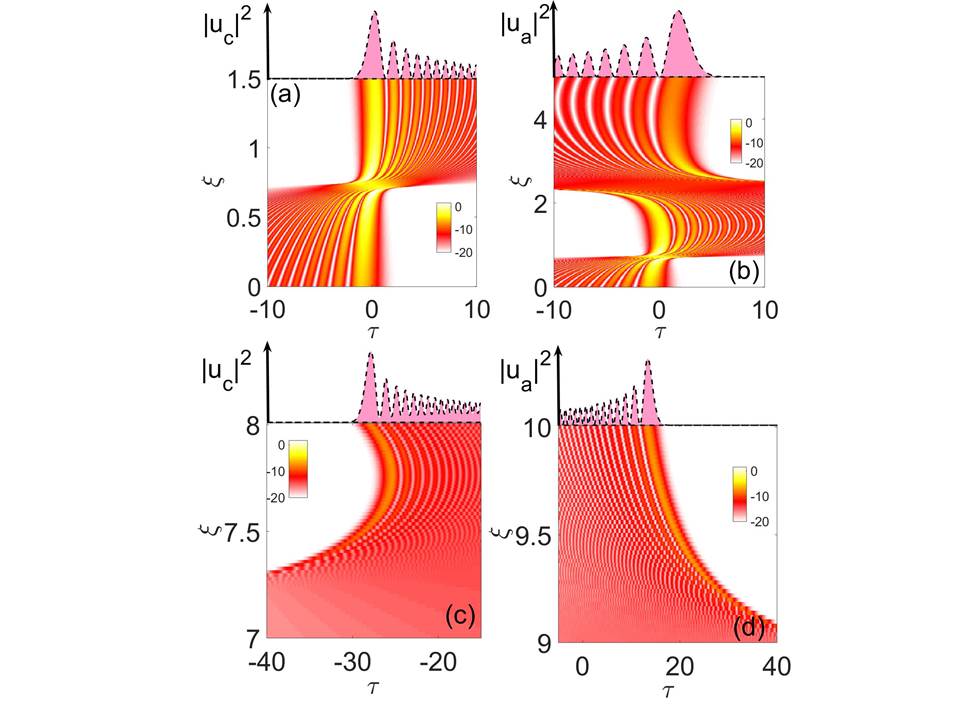}
  \vspace{-1em}
 \caption{Density plots of different zones for $\chi_3=0.5$. The different zones are represented by different integer values in Eq.\ref{q9}. The analytical solutions(black dashed lines) obtained in Eq.\ref{q8}(for(b) and (d)  and Eq.\ref{q13}(for (a)and (c)(pink shaded area) are compared in the upper panels of each figure.  }
 \label{Figure10}
\end{center}
 \end{figure}

\begin{figure}[h!]
 \begin{center}
  \includegraphics[trim=0.0in 1.1in 0.1in 0.7in,clip=true,  width=83mm]{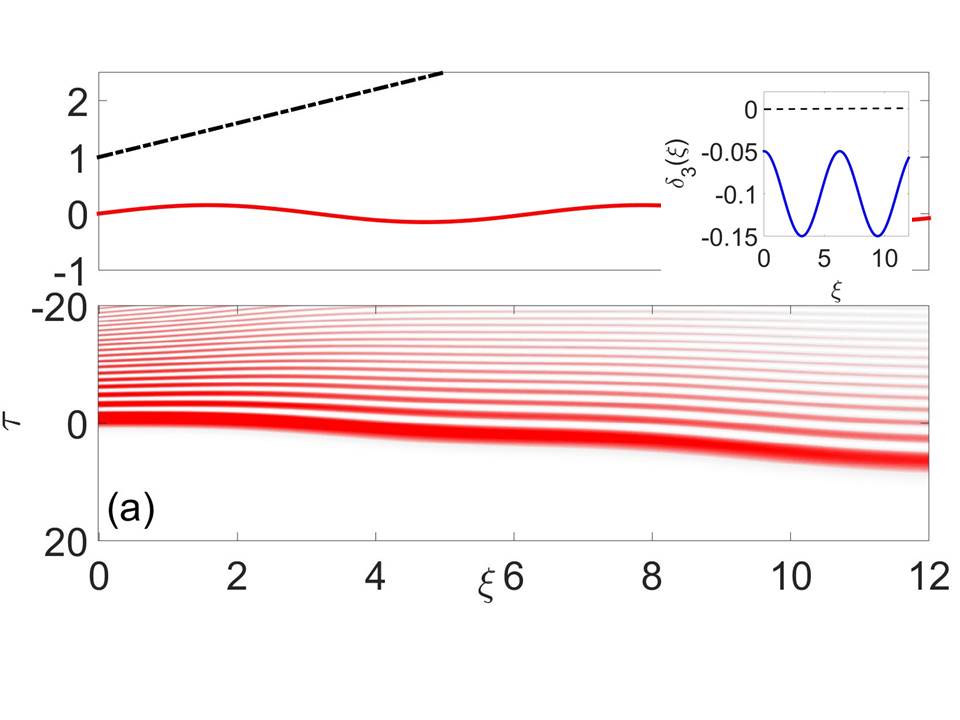}
  \includegraphics[trim=0.1in 1.0in 0.1in 0.5in,clip=true,  width=83mm]{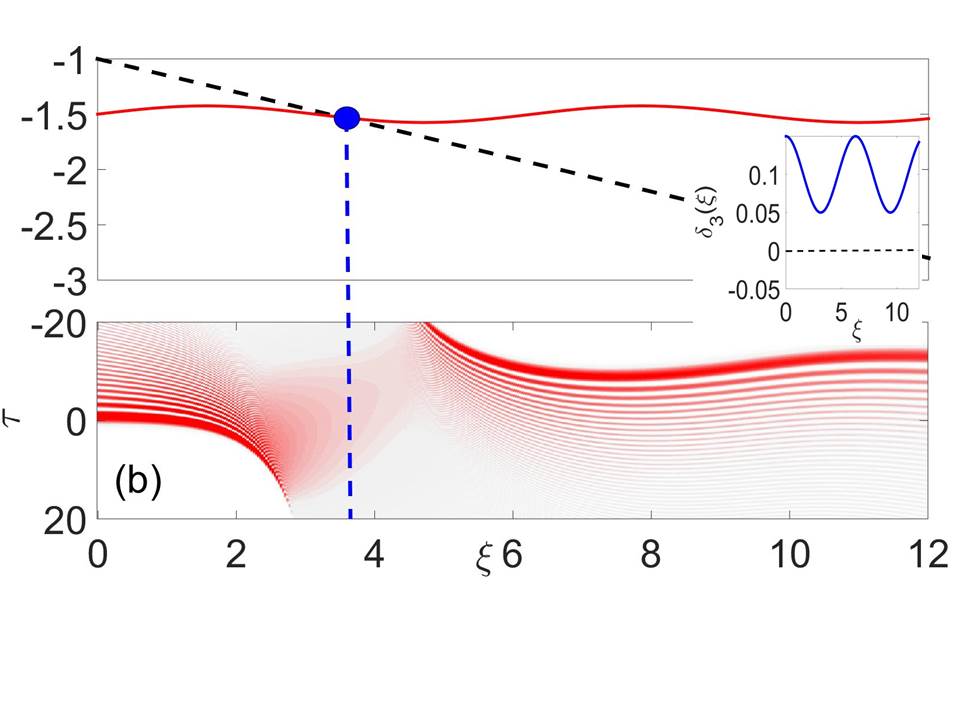}
  \includegraphics[trim=0.0in 1.0in 0.2in 0.7in,clip=true,  width=83mm]{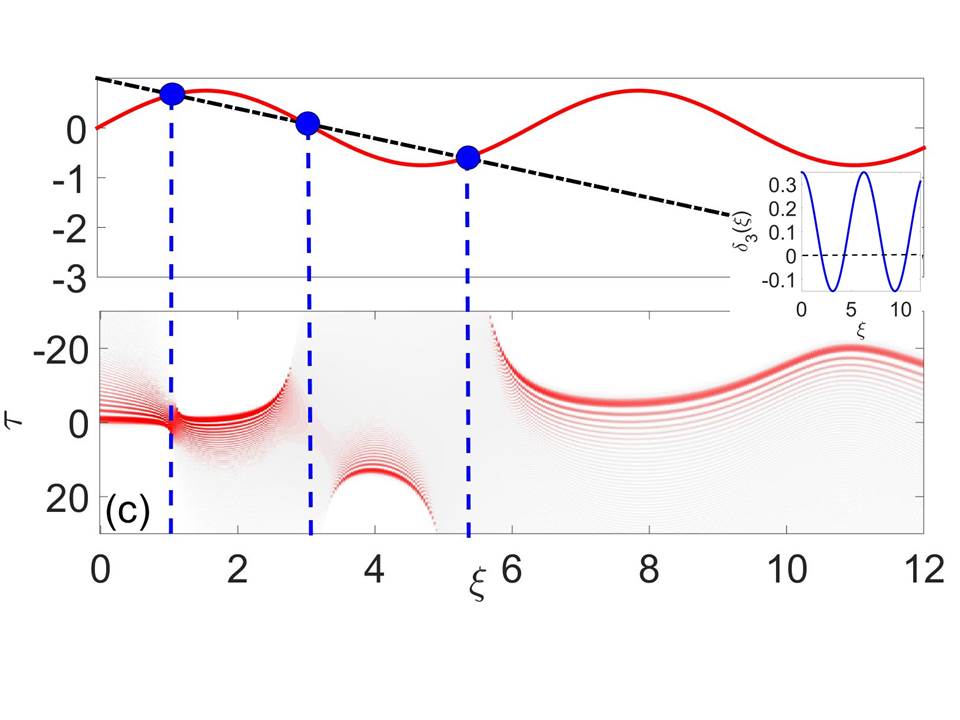}
  \vspace{-1em}
 \caption{The dynamics of the Airy pulse under different conditions (a) $\bar{\delta}_{30}=-0.1; \chi_3=0.05$ (b) $\bar{\delta}_{30}=0.1;\chi_3=0.05$ (c) $\bar{\delta}_{30}=0.1;\chi_3=0.25$. The variation of $\delta3$ is provided in the insets. It can be seen that the flipping conditions of the pulse depends on the geometrical variation of $\delta_3$ and its initial value. In each case the flipping condition arises  when the transcendental equation (Eq.\eqref{eq15}) have real solution as indicated in the figure.    }
                 
\label{Figure11}
\end{center}
 \end{figure}

It should be noted that the value of $\xi_{fj}^{(n)}$ depends on the integers $n$ (see Eq.\eqref{q9}) which determines the position of the flipping zone through $\chi_3$. The selective focusing of a particular zone (say $n^{th}$ zone) will be achieved for particular $\chi$ determined by the Eq.(\ref{q12}). We illustrate this complex phenomena graphically in Fig.\ref{Figure9} through the density plot. We can notice that the flipping zone can be merged to a point (marked by black circle) selectively. Here we take the first flipping position ($n=0$) in Eq.\eqref{q9} for $j=1$ and use this value to find the parameter $\chi_f$  for which the tight focusing happens at $\xi_{f1}^{(0)}$ (Fig.\ref{Figure9}(a)) . Similarly for the second position(Fig.\ref{Figure9}(b)) we consider $\xi_{f2}^{(0)}$  in Eq.\ref{q9} with $n=0$ and $j=2$ and calculate $\chi_f$ from Eq.\ref{q12} . Using this technique we can selectively focus the FEAP according to our requirement. We can see that the presence of periodic TOD complicates the dynamics of the FEAP and divide its propagation into many distinct zones. Under static TOD the propagating FEAP experiences a singularity and we can divide the entire propagating length into three distinct zones (i) zone-I before flipping (ii) zone-II flipping and (iii) zone-III after flipping. In zone-I the airy pulse moves with its usual ballistic trajectory. In zone-II the pulse experiences a singularity and try to confine in a finite region. In zone-III the airy pulse temporally flips. When TOD is periodic over distance then, the FEAP flips over periodically against each focusing. For detail investigation we require the analytical solution of the FEAP at each zone.  In Eq.\ref{q8} we obtain the solution of the propagating FEAP under periodic TOD. This solution works well in zone-I and valid in the zones after the odd numbered flipping ($2^{nd}$,$4^{th}$ etc) positions which are achieved for second set of flipping points $\xi_{f2}^{(n)}$ in Eq.\ref{q9}. We also derive the solution at flipping position (zone-II) where FEAP completely loses its characteristics and converted to a pure Gaussian pulse as given in   Eq.\ref{q10}. The solutions of the pulse for the zones beyond the first set of flipping point ($\xi_{f1}^{(n)}$ in Eq.\ref{q9} can be obtained by defining a variable transformation $\xi'=\xi-\xi_{f1}^{(n)}$.  Under this new variable the solution can be expressed as 

\begin{equation}\label{q13}
\begin{aligned}
 u_c(\xi',\tau)=\frac{1}{c'}\exp\left({\frac{a^3}{3}-\widetilde{\alpha}\xi}\right)Ai\left(\frac{b'}{c'}-\frac{n'^2}{c'^4}\right)\\
 \exp i\left(\frac{2n'^3}{3c'^6}-\frac{n'b'}{c'^3} \right),
     \end{aligned} 
     \end{equation}  
 where $c'=[3\chi_3(\sin\xi-\sin\xi_{f1}^{(n)})]^\frac{1}{3}$, $b'=-\tau$ and $n'=ia-\Delta^{(n)}-\frac{\xi'}{2}+\frac{\chi}{2}[\sin(\xi)-\sin(\xi_{f1}^{(n)})]$. In Fig.\ref{Figure10} we illustrate the dynamics of FEAP as shown in Fig.\ref{Figure7}(a) zone wise. The aim here is to check the validity of the analytical solutions that we obtain in Eq.\ref{q8} and Eq.\ref{q13}. In Fig.\ref{Figure10}(a) we demonstrate the dynamics of FEAP experiencing first flipping at $\xi=\xi_{f1}^{(n=0)}$ where as in Fig.\ref{Figure10}(b) the pulse move forward and encounter the next singularity at $\xi=\xi_{f2} ^{(n=0)}$. For both cases the derived analytical solution (dashed lines) corroborate well with numerical output (pink shaded area). We extended out investigation for $\xi>\xi_{f1}^{(n=1)}$ (Fig.\ref{Figure10}(c)) and $\xi>\xi_{f2}^{(n=1)}$ (Fig.\ref{Figure10}(d)) and  find good agreement with numerical and theoretical results.

Finally we conclude our work by investigating the airy pulse dynamics for a general $\delta_3 (\xi)$ variation where $\bar{\delta}_{30} \neq0$. In such case we do not expect any periodic focusing. The general transcendental equation that governs the focusing is, 
\begin{equation}\label{eq15}
\sin(\mu\xi)=\frac{\mu}{3 \chi_3}(1-3\bar{\delta}_{30} \xi)
\end{equation}
Note, when $\delta_3(\xi)<0$ we must have $\bar{\delta}_{30} <0$ and $|\bar{\delta}_{30}|/\chi_3 >1$. It is easy to show that if $|\bar{\delta}_{30}|/\chi_3 >1$ we do not have any solution of Eq. (\eqref{eq15}). In other word, when TOD coefficient is throughout negative ($\delta_3(\xi)<0$) there will be no temporal focusing of FEAP. To illustrate this feature in Fig. \ref{Figure11}(a) we plot the dynamics of a FEAP under modulating TOD coefficient which is throughout negative. The situation is different when $\delta_3(\xi) >0$ where we can have only one solution of Eq. \eqref{eq15} defining the flipping of FEAP. In Fig. \ref{Figure11}(b)  we demonstrate the flipping of the airy pulse at the precise location ($\xi_c$) where $\xi_c$ satisfies Eq. \eqref{eq15}. However more than one flipping can possible when numeric sign of $\delta_3$ varies from positive to negative values. In such case multiple solution of the Eq. \eqref{eq15} is possible and each solution defines the flipping as illustrated in Fig. \ref{Figure11}(c).

\section {Conclusion}
In this report we investigate the dynamics of a finite energy Airy pulse (FEAP) under the environment of varying dispersion. We propose realistic waveguide structure that offer linear and oscillatory GVD profile as a function of propagation distance.  A detail analysis reveals linear variation of GVD affects the usual ballistic trajectory of an Airy pulse. By suitably adjusting the modulation strength parameter one  can even achieve an unusual quasi-linear trajectory for FEAP.   It is also found that the  power  carried by the primary lobe can be manipulated for varying dispersion parameters. We theoretically estimate a critical value of the modulation strength of the varying GVD parameter for which the power attenuation of the main lobe is minimal.  Our theoretical results agrees well with numerical simulation. The dynamics of FEAP is found  to be very interesting under oscillatory second and third order dispersion (TOD). The presence of oscillatory TOD offers multiple singularity zones where wings of the Airy pulse flips temporally. The dynamics of an Airy pulse near singular points is very rich and demands special investigation. We meticulously solve the propagation equation and identify the location of flipping position in ($\xi-\tau$) plane. The theoretical calculation reveals the physical condition of getting selective absolute focusing where peak power of the propagating Airy pulse reaches to its maxima. All the analytical findings are supported by adequate numerical simulation throughout the report.  The manipulation of Airy trajectory and power level using the concept of varying dispersion might be useful in practical purposes.

\section*{Acknowledgements}
A.B. acknowledges Ministry of Human Resource Development (MHRD), India for a research fellowship.

\end{document}